 \documentclass[final,5p,times,twocolumn]{elsarticle}

\usepackage{amssymb}
\usepackage{lipsum}
\usepackage{braket}
\usepackage{url}
\usepackage{hyperref}
\usepackage[T1]{fontenc}
\usepackage{amsmath}

\journal{Physics Letters B}

\begin{document}

\begin{frontmatter}



\title{Coulomb Excitation of $^{80}$Sr and the limits of the $N=Z=40$ island of deformation}



\author[surrey]{\mbox{R. Russell}}      

\author[surrey]{\mbox{J. Heery}}

\author[surrey]{\mbox{J. Henderson}\corref{email}}
    \ead{jack.henderson@surrey.ac.uk}

\author[york]{\mbox{R. Wadsworth}}
    
\author[sangyo]{\mbox{K. Kaneko}}
    
\author[tsukuba]{\mbox{N. Shimizu}}
    
\author[senshu]{\mbox{T. Mizusaki}}
    
\author[shanghai]{\mbox{Y. Sun}}

\author[sfu_C]{\mbox{C. Andreoiu}}

\author[sfu_C]{\mbox{D. W. Annen}}

\author[triumf]{\mbox{A. A. Avaa}}  

\author[triumf]{\mbox{G. C. Ball}}

\author[guelph]{\mbox{V. Bildstein}}

\author[guelph]{\mbox{S. Buck}}

\author[surrey]{\mbox{C. Cousins}}

\author[triumf]{\mbox{A. B. Garnsworthy}}

\author[frib]{\mbox{S. A. Gillespie}}

\author[guelph]{\mbox{B. Greaves}}

\author[triumf]{\mbox{A. Grimes}}

\author[triumf]{\mbox{G. Hackman}}

\author[llnl]{\mbox{R. O. Hughes}}

\author[york]{\mbox{D. G. Jenkins}}

\author[unc,tunl]{\mbox{T. M. Kowalewski}} 

\author[sfu_P]{\mbox{M. S. Martin}}

\author[anl]{\mbox{C. M\"{u}ller-Gatermann}}

\author[triumf]{\mbox{J. R. Murias}}

\author[triumf]{\mbox{S. Murillo-Morales}}


\author[surrey,bucharest]{\mbox{S. Pascu}}

\author[triumf,llnl]{\mbox{D. M. Rhodes}}

\author[jaea]{\mbox{J. Smallcombe}}

\author[sfu_C]{\mbox{P. Spagnoletti}}

\author[triumf,guelph]{\mbox{C. E. Svensson}}

\author[york]{\mbox{B. Wallis}}

\author[triumf]{\mbox{J. Williams}}

\author[llnl]{\mbox{C. Y. Wu}}

\author[triumf,ubc]{\mbox{D. Yates}}

\cortext[email]{Corresponding author:}

\affiliation[surrey]{addressline={School of Maths and Physics, University of Surrey, Guildford, GU2 7XH, Surrey}, country={United Kingdom}}

\affiliation[york]{addressline={School of Physics, Engineering and Technology, University of York, York, YO10 5DD, North Yorkshire}, country={United Kingdom}}

\affiliation[sangyo]{addressline={Department of Physics, Kyushu Sangyo University, Fukuoka 813-8503}, country={Japan}}

\affiliation[tsukuba]{addressline={Center for Computational Sciences, University of Tsukuba, Tennodai, Tsukuba 305-8577}, country={Japan}}

\affiliation[senshu]{addressline={Institute of Natural Sciences, Senshu University, Tokyo 101-8425}, country={Japan}}

\affiliation[shanghai]{addressline={School of Physics and Astronomy, Shanghai Jiao Tong University, Shanghai 200240}, country={China}}

\affiliation[sfu_P]{addressline={Department of Physics, Simon Fraser University, Burnaby, V5A 1S6, British Columbia}, country={Canada}}
\affiliation[sfu_C]{addressline={Department of Chemistry, Simon Fraser University, Burnaby, V5A 1S6, British Columbia}, country={Canada}}

\affiliation[triumf]{addressline={TRIUMF, Vancouver, V6T 2A3, British Columbia}, country={Canada}}

\affiliation[guelph]{addressline={Department of Physics, University of Guelph, Guelph, N1G 2W1, Ontario}, country={Canada}}

\affiliation[frib]{addressline={Facility for Rare Isotope Beams, Michigan State University, East Lansing, 48824, Michigan}, country={USA}}

\affiliation[llnl]{addressline={Lawrence Livermore National Laboratory, Livermore, 94550, California}, country={USA}}

\affiliation[unc]{addressline={Department of Physics and Astronomy, University of North Carolina, Chapel Hill, 27599, North Carolina}, country={USA}}
\affiliation[tunl]{addressline={Triangle Universities Nuclear Laboratory, Duke University, Durham, 27708, North Carolina}, country={USA}}

\affiliation[anl]{addressline={Physics Division, Argonne National Laboratory, Lemont, 60439, Illinois}, country={USA}}


\affiliation[bucharest]{addressline={National Institute for Physics and Nuclear Engineering , Bucharest-Magurele, R-77125}, country={Romania}}

\affiliation[jaea]{addressline={Japan Atomic Energy Agency, Naka-gun, 319-1184, Ibaraki}, country={Japan}} 

\affiliation[ubc]{addressline={Department of Physics and Astronomy, University of British Colombia, Vancouver, V6T 1Z4, British Colombia}, country={Canada}}


\begin{abstract}
The region of $N\approx Z\approx 40$ has long been associated with strongly deformed nuclear configurations. The presence of this strong deformation was recently confirmed through lifetime measurements in $N\approx Z$ Sr and Zr nuclei. Theoretically, however, these nuclei present a challenge due to the vast valence space required to incorporate all deformation driving interactions. Recent state-of-the-art predictions indicate a near axial prolate deformation for $N=Z$ and $N=Z+2$ nuclei between $N=Z=36$ and $N=Z=40$. In this work we investigate the shores of this island of deformation through a sub-barrier Coulomb excitation study of the $N=Z+4$ nucleus, \textsuperscript{80}Sr. Extracting a spectroscopic quadrupole moment of $Q_s(2^+_1) = 0.45^{+0.83}_{-0.88}$~eb, we find that \textsuperscript{80}Sr is inconsistent with significant axial prolate deformation. This indicates that the predicted region of strong prolate deformation around $N=Z=40$ is tightly constrained to the quartet of nuclei: \textsuperscript{76,78}Sr and \textsuperscript{78,80}Zr.
\end{abstract}


\end{frontmatter}


\section{Introduction}\label{introduction}

The atomic nucleus exhibits a number of emergent properties arising from the underlying structure of its nucleons. Among these is collectivity, arising from the collective motion of nucleons as the nucleus deviates from sphericity, which is a key test of nuclear models, necessarily involving multiple nucleons within the nucleus. Quadrupole deformation in the upper $fpg$ region of nuclei, delimited by the $N=Z$ nuclei \textsuperscript{56}Ni and \textsuperscript{100}Sn, has long challenged configuration-interaction type nuclear models. This is due to the strong deformation driving effect of the quadrupole-quadrupole ($QQ$) interaction between the $g_{9/2}$ orbital below $N=Z=50$ and the $2d_{5/2}$ orbital residing outside of the nominal $fpg$ space, requiring a vast valence space to capture the necessary correlations. This deformation-driving effect is further exacerbated at the line of $N=Z$, with the protons and neutrons experiencing enhanced correlations due to their large spatial overlap.

The region around $N=Z=38$ \textsuperscript{76}Sr and $N=Z=40$ \textsuperscript{80}Zr has long been associated with strong deformation~\cite{ref:Lister_82,ref:Lister_87,ref:Sahu_88}. The extent of the quadrupole deformation in the region was recently verified through lifetime measurements~\cite{ref:Llewellyn_20} of the first $2^+$ state in \textsuperscript{80}Zr. This work also confirmed the enhanced deformation previously reported in \textsuperscript{76}Sr~\cite{ref:Lemasson_12}. These measurements yield $B(E2)=120(^{+18}_{-14})$ W.u. and $B(E2)=93(9)$ W.u. for \textsuperscript{76}Sr and \textsuperscript{80}Zr, respectively, indicative of strongly deformed ($\beta_2\approx0.4$) nuclear systems. In contrast to this strong deformation however, forty is a harmonic oscillator magic number, at which the $3\hbar\omega$ shell would nominally be filled. While the strong spin-orbit interaction in the atomic nucleus breaks down this shell closure, its effects might be expected to linger somewhat, especially given the doubly-magic nature of \textsuperscript{90}Zr~\cite{ref:Garrett_03,ref:Kaneko_06}, only ten neutrons away. Perhaps unsurprisingly, therefore, \textsuperscript{80}Zr has previously been predicted to exhibit multiple shape coexistence, with five $0^+$ states predicted below 2.25~MeV~\cite{ref:Rodriguez_11}. Recent developments in shell-model theory have yielded promising results in this region, demonstrating the emergence of deformation at $N=Z=40$ using the PMMU interaction~\cite{Kaneko21}. In that work, $N=Z$ and $N=Z+2$ nuclei were investigated, with collective observables extracted and the role of the $gd$ $QQ$ interaction tracked. It was found that calculations incorporating the $gd$ interaction predicted strongly deformed, prolate systems between $N=Z=36$ and $N=Z=40$, whereas the suppression of the $gd$, $QQ$ interaction entirely eliminated this effect, resulting in modestly deformed oblate systems. 

In this letter we present the first sub-barrier Coulomb excitation measurement of \textsuperscript{80}Sr, lying only two protons and two neutrons ``South East'' of $N=Z=40$ \textsuperscript{80}Zr. In doing so we are able to access for the first time the spectroscopic quadrupole moment of the first $2^+$ state, $Q_s(2^+_1)$. This is a key observable in understanding the collective properties of the nucleus, and can be related to the form of the underlying nuclear deformation: prolate vs oblate vs triaxial. We are therefore able to determine, for the first time, the limits of the long-predicted island of strong, axial, prolate deformation around $N=Z=40$.

\section{Experimental Methods}\label{Experimental Methods}

The experiment was carried out at the TRIUMF-ISAC facility, with the use of the $\gamma$-ray detection array TIGRESS~\cite{Hackman14, Scraggs05}. Radioactive \textsuperscript{80}Sr was produced using the isotope-separation online (ISOL) method. The TRIUMF cyclotron impinged a 520-MeV proton beam upon a thick niobium target, producing radioisotopes which were then extracted using the Ion-Guide Laser Ion Source (IG-LIS)~\cite{Raeder14}. IG-LIS suppresses surface-ionised products using a repeller electrode held at 40~V, permitting only neutral atoms to escape the target volume. After diffusing from the target volume, the strontium atoms are selectively ionised through resonant ionisation and mass separated, providing a purified beam of \textsuperscript{80}Sr. The singly-charged ions are then charge bred using the electron cyclotron resonance (ECR) charge state breeder (CSB)~\cite{Jayamanna02, Ames05}, prior to injection into the ISAC and ISAC-II accelerator chain. The use of the CSB introduces stable contaminants into the beam, of which the most problematic for the present work are isobars of \textsuperscript{80}Sr, in particular \textsuperscript{80}Se which dominated the beam composition, and \textsuperscript{80}Kr. The beam was then accelerated to $4.25 \,$MeV/u and delivered to the experimental station. 

\begin{figure}[t] 
\centering
\includegraphics[width=\columnwidth]{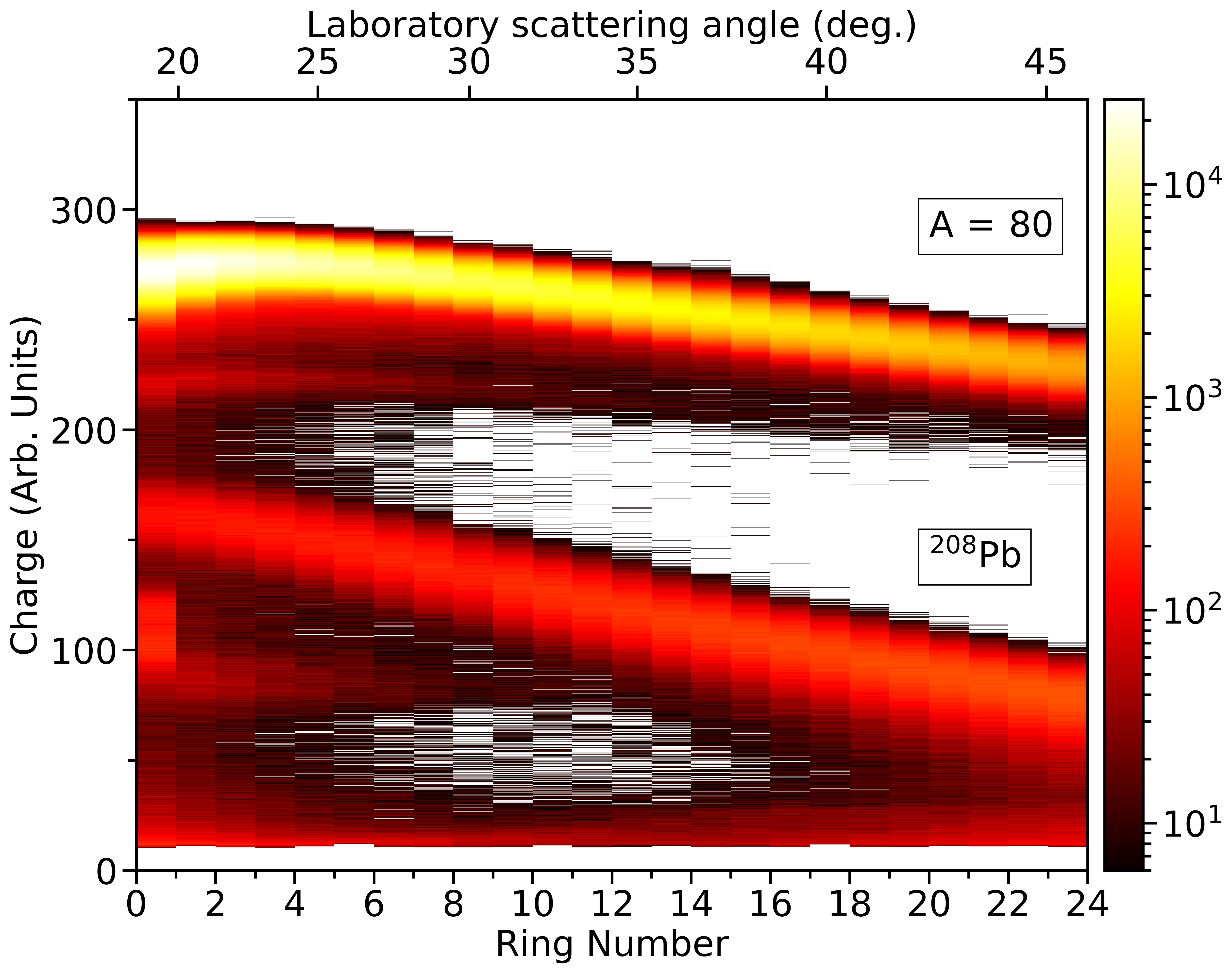}
\caption{Charge detected in the rings of the downstream S3 detector. The laboratory angles are shown on the upper axis. The top band are the A=80 nuclei, such as the desired $^{80}$Sr, and the lower band are the $^{208}$Pb nuclei which have been scattered by the beam-like particles. Other smaller features in the data are attributed to $A/Q$ contaminants in the beam, present at a lower level than the $A=80$ nuclei and excluded on the basis of kinematic selection. Note that, for the purpose of presentation, bins containing less than 5 counts have been omitted from this figure.}\label{Charge-RingNum}
\end{figure}

\begin{figure}[t]
    \centering
    \includegraphics[width = \columnwidth]{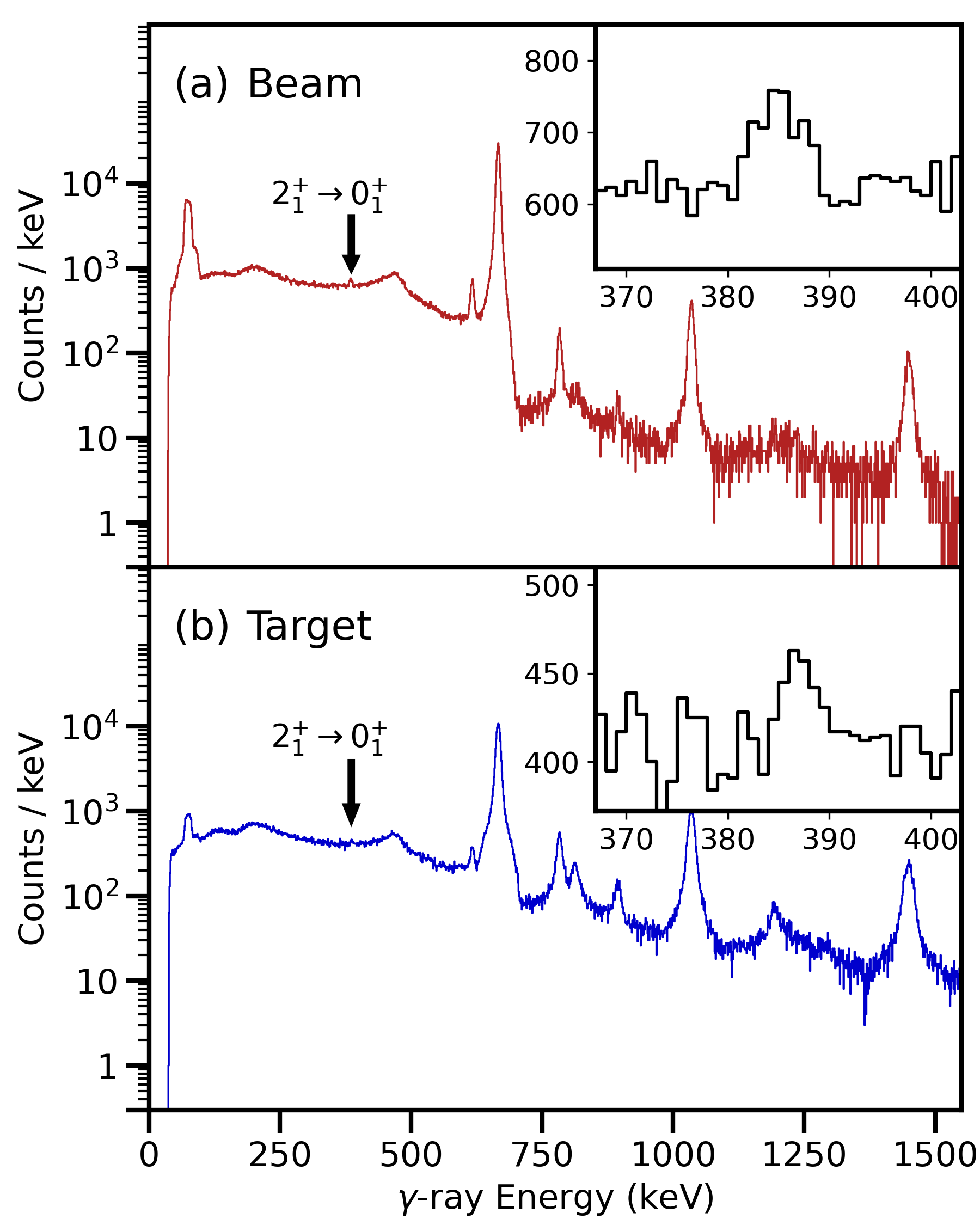}
    \caption{Addback $\gamma$-ray spectra with Compton suppression for detected A=80 nuclei with a Doppler correction for A=80 nuclei (a) and for detected $^{208}$Pb nuclei with a Doppler correction for A=80 nuclei (b) shown. Insets: These show the 386$\,$keV $2^{+}_{1} \rightarrow 0^{+}_{1}$ transition for $^{80}$Sr. The unmarked major peaks seen in this figure are from other A=80 beam-like projectiles.}
    \label{fig:Doppler_Spectra}
\end{figure}

The beam was impinged upon a $0.882 \,$mg/cm$^{2}$ target of $^{208}$Pb at the centre of TIGRESS, with S3-type annular silicon particle detectors located up- and down-stream of the target. Excited states in the beam and target nuclei were populated through Coulomb excitation, with scattered nuclei detected in the S3 detectors. Prompt $\gamma$-rays from the de-excitation of the populated excited states were detected in TIGRESS. The data were analysed using the GRSISort~\cite{GRSISort} analysis package built in the ROOT~\cite{ROOT} framework. Scattered particles were selected and identified as $A=80$ (beam-like) and $A=208$ (target-like) on the basis of their kinematics, as shown in Fig.~\ref{Charge-RingNum}. Gamma-ray energies were added-back based on coincident hits in immediately neighbouring HPGe crystals to maximise efficiency and suppress Compton background. Compton-scatters were further suppressed on the basis of hits in the TIGRESS Compton veto detectors. Gamma-ray energies were then corrected for their Doppler shift on the basis of the two-body scattering kinematics, yielding the $\gamma$-ray spectra shown in Fig.~\ref{fig:Doppler_Spectra}. Clearly, the spectra are dominated by the excitation of the strong $^{80}$Se contaminant in the beam, with a $2^+_1\rightarrow0^+_1$ transition energy of 666~keV. The $2^+_1\rightarrow0^+_1$ transition in \textsuperscript{80}Sr was, however, visible above the Compton scattering background at 386~keV~(see Fig.\ref{fig:Doppler_Spectra}), allowing for its yield as a function of scattering angle to be extracted.

Yields from Coulomb excitation were evaluated using the semi-classical coupled channels Coulomb excitation code GOSIA~\cite{GOSIA}, with the SRIM~\cite{SRIM} software package employed to account for energy loss of ions in the target. An external $\chi ^{2}$ minimisation was used, in conjunction with GOSIA, which uses the ROOT MINUIT libraries~\cite{GOSIA_Fitter}. The experimental data were divided into six angular ranges for the downstream detector for each kinematic solution (beam- and target-like ion detection) and a single angular range for the upstream $A=80$ detection, for a total of thirteen angular ranges. This provides for a near continuous coverage from approximately $30^\circ\rightarrow160^\circ$ in the centre-of-mass frame. Due to the large \textsuperscript{80}Se contamination, normalisation to a target transition was not possible and the minimisation therefore used the adopted literature value of B(E2; $2^{+}_1 \rightarrow 0^{+}_1$) as an additional constraint~\cite{NNDC}. 

\section{Results}\label{Results} 

The dominant source of uncertainty in the present result arose from the large Compton background on which the 386-keV \textsuperscript{80}Sr peak was located. Every effort was made to reduce this background, including the use of the Compton suppression shields mounted on the TIGRESS array and the application of appropriate multiplicity gates to the experimental data. Even with these efforts, the uncertainty remained dominated by this background, to the extent that all other systematic contributions to the uncertainty are largely negligible. Nonetheless, we briefly discuss some systematic contributions below that were included within the analysis.

One key contribution to Coulomb excitation is the role of multi-step excitations. These excitations can result in interference terms due to the relative signs of the matrix elements involved. Of particular importance are the relative signs of $E2$ matrix elements between $2^+$ states. To understand any additional contribution of these interference terms, signs and magnitudes of $E2$ matrix elements connecting to the $2^+_2$ and $2^+_3$ were varied. It was found that the impact of this variation was considerably smaller than the statistical uncertainty obtained in the $\chi^2$ minimisation, however its variance was obtained and is included in the quoted result.

An additional consideration in this analysis was the low $0^{+}_2$ state observed in Ref.~\cite{Alford79} to be at $1000 (100)\,$keV. This level is expected to have an E2 transition to $2^{+}_1$ and an E0 transition to $0^{+}_1$ similar to those seen in work for other nuclei in this region~\cite{Giannatiempo95, Giannatiempo86}. To ensure that there was no undue strength attributed to this $0^{+}_2 \rightarrow 2^{+}_1$ transition, an E0 transition will need to be accounted for in the analysis for which GOSIA has no facility at present. This leads to the method seen to be used in Refs.~\cite{Wrozsek-Lipska2019, Kesteloot15, Clement16} where a pseudo $1^{+}$ level is made in GOSIA for the purpose of adding M1 de-excitations following the $0^{+}_2 \rightarrow 1^{+}_{pseudo} \rightarrow 0^{+}_1$ path. It was found that there was no strong influence of this transition on the experimental results obtained here, with a systematic contribution to $Q_s(2^+_1)$ of 0.01~eb.

\begin{figure}
    \centering
    \includegraphics[width=\linewidth]{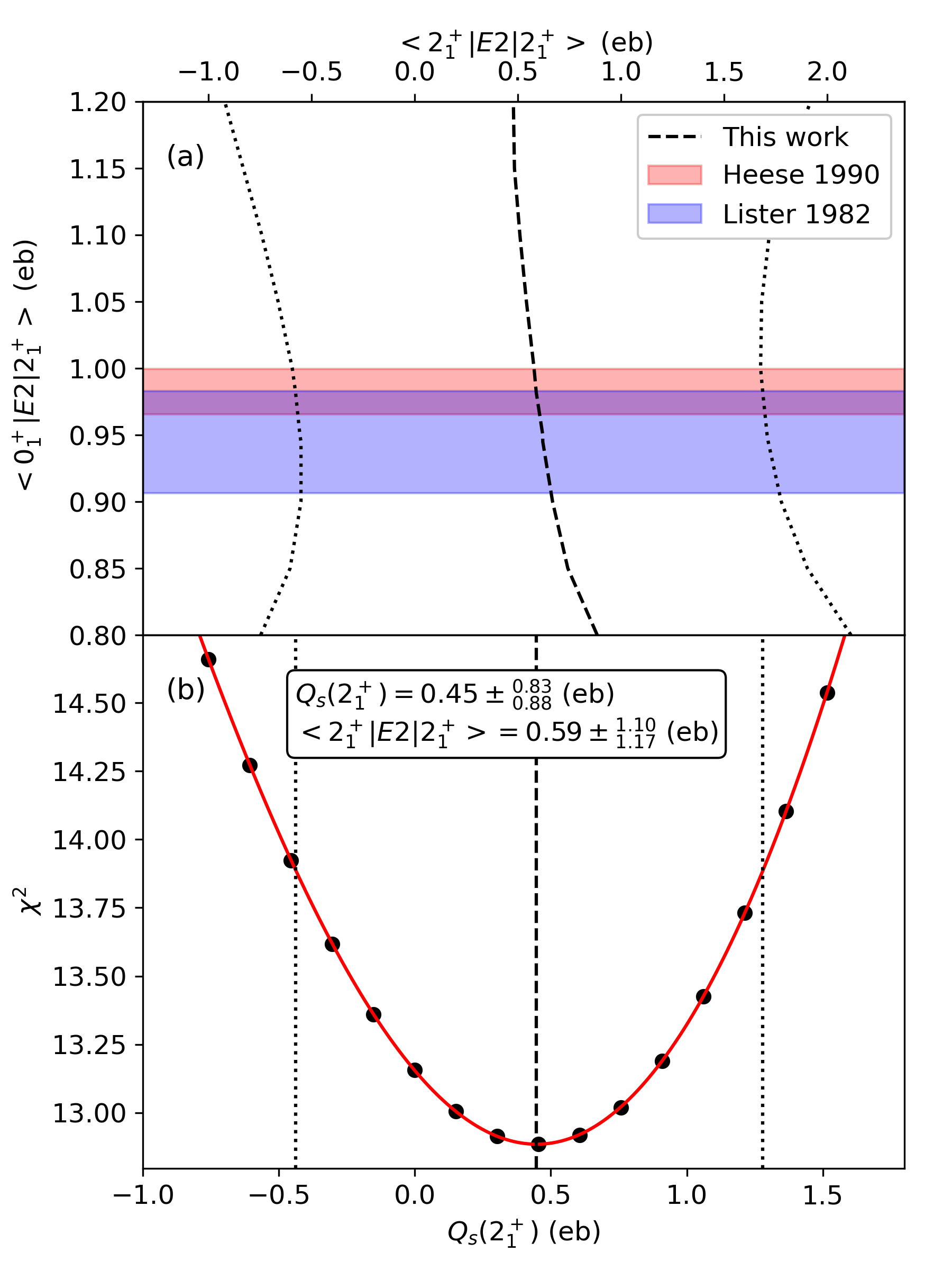}
    \caption{(a) $Q_s(2^+_1)$ extracted in the present work, plotted against the $\bra{0^+_1}E2\ket{2^+_1}$ matrix element. The bottom axis is $Q_s(2^+_1)$ while the top axis is the corresponding matrix element $\bra{2^+_1}E2\ket{2^+_1}$. The quoted $Q_s(2^+_1)$ in this work is uses the result from~\cite{ref:Heese_90} (red band), which is taken as the evaluated~\cite{NNDC} value. The intersection of the present data with the second most precise result, taken from Ref.~\cite{ref:Lister_82} is also shown, having only a very minor impact on the final extracted $Q_s(2^+_1)$ value. (b) The fitted $\chi^2$ plotted as a function of $\bra{2^+_1}E2\ket{2^+_1}$ matrix element (top axis) and $Q_s(2^+_1)$ (bottom axis). Shown by the dashed line is the minimum value, while the dotted lines indicate the $\chi^2_{min}+1$ limits, yielding $Q_s(2^+_1)=0.45^{+0.83}_{-0.88}$~eb.}
    \label{fig:Chisq}
\end{figure}

With the above considerations and analysis, we obtain $Q_s(2^+_1) = 0.45^{+0.83}_{-0.88}$~eb in $^{80}$Sr. The $\chi^2$ distribution used to extract this value and uncertainties is shown in Fig.~\ref{fig:Chisq}. While the present result has large uncertainty, it still excludes a large range of the available $Q_s(2^+_1)$ space permitted within the bounds of standard collective behaviour. To demonstrate this, we consider $Q_s(2^+_1)/\left|Q^{rot}_s(2^+_1)\right|$, where 

\begin{equation}
    Q^{rot}_s(2^+_1) = {\frac{2}{7}\sqrt{\frac{16\pi}{5} \cdot B(E2;0^+_1\rightarrow2^+_1)}}. 
    \label{eq:Qrot}
\end{equation}

This ratio, also referred to in the literature as the reduced quadrupole moment ($q_s$)~\cite{ref:Rhodes_21}, can also be equated approximately to $-\cos(3\gamma)$ in the strongly-deformed regime~\cite{ref:Henderson_20}. The experimental value for $Q_s(2^+_1)$ yields $Q_s(2^+_1)/\left|Q^{rot}_s(2^+_1)\right|=0.5(9)$, which is inconsistent with the prolate limit ($Q_s(2^+_1)/\left|Q^{rot}_s(2^+_1)\right|=-1$) at the level of more than $1.5\sigma$.

\begin{figure}
    \centering
    \includegraphics[width=\linewidth]{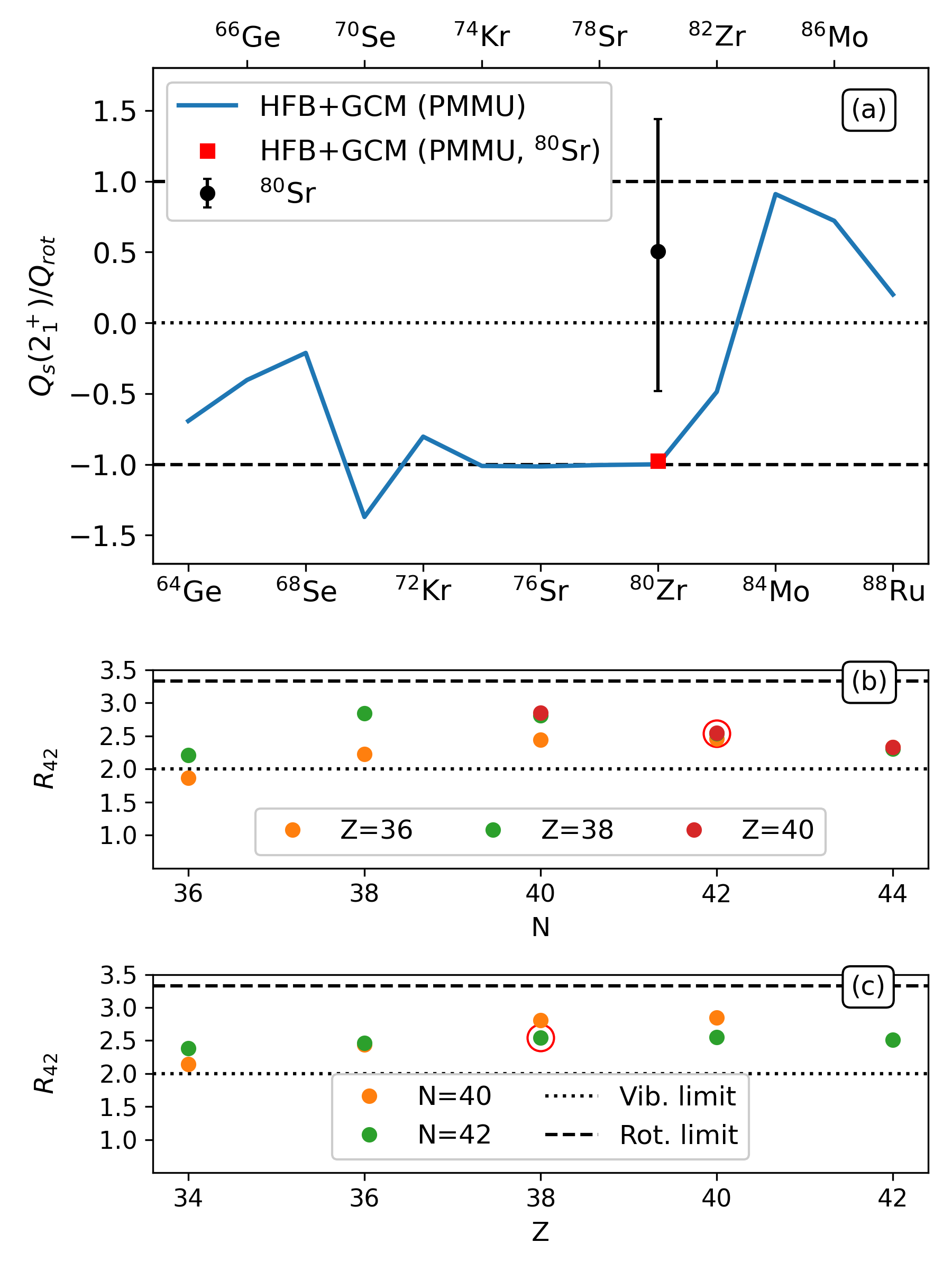}
    \caption{(a) The ratio of the spectroscopic quadrupole moment ($Q_s(2^+_1)$) to the rotational limit ($Q_{rot}$) defined in Eq.~\ref{eq:Qrot}, as calculated from Ref.~\cite{Kaneko21} using the PMMU interaction with the Hartree-Fock Bogoliubov plus Generator Coordinate Method (HFB + GCM) method. Also shown is the experimental ratio deduced in the present work for \textsuperscript{80}Sr, and the theoretical value for \textsuperscript{80}Sr, calculated using the HFB + GCM method (red square). (b) The ratios of the $4^+_1$ energy to the $2^+_1$ energy ($R_{42}$) for Kr ($Z=36$), Sr ($Z=38$) and Zr ($Z=40$) isotopes as a function of neutron number, $N$. (c) $R_{42}$ for $N=40$ and $N=42$ isotones, as a function of proton number, $Z$. Strontium-80 is highlighted in both (b) and (c) with a red circle. The vibrational and rotational $R_{42}$ limits are indicated by the dotted and dashed lines, respectively. }
    \label{fig:PMMU_Calculations}
\end{figure}

We now compare this result to state-of-the art calculations performed using the method outlined in Ref.~\cite{Kaneko21}. Here, the PMMU interaction is used with a large model space ($2p_{3/2},f_{5/2},2p_{1/2},1g_{9/2},2d_{5/2}$). This model space has been shown to be essential in reproducing prolate deformation in $N=Z$ nuclei beyond $A=70$, where the quadrupole-quadrupole $gd$ interaction plays a dominant role in driving the nuclear deformation. Further to results for \textsuperscript{76,78}Sr and other $N=Z$ and $N=Z+2$ isotopes, presented in Ref.~\cite{Kaneko21}, we present here calculations for \textsuperscript{80}Sr, summarised in Fig.~\ref{fig:PMMU_Calculations} (a). The PMMU calculations predict a large, negative $Q_s(2^+_1)$ value, at approximately the prolate axial limit (i.e. $Q_s(2^+_1)/\left|Q^{rot}_s(2^+_1)\right|\approx-1$), in disagreement with the present experimental result. We note, however, that the calculations of Ref.~\cite{Kaneko21} predict a shape transition to occur between $A=80$ and $A=84$, along the line of $N=Z$, with prolate and oblate shape-minima near-degenerate at \textsuperscript{84}Mo. In addition, whereas PMMU shell model calculations predict a sudden transition from axial prolate to axial oblate deformation in Ref.~\cite{Kaneko21}, HFB+GCM calculations predict an intermediate nucleus: \textsuperscript{82}Zr, in which $Q_s(2^+_1)/\left|Q^{rot}_s(2^+_1)\right|\approx-0.5$ lies between the two axial solutions. One might therefore hypothesise that \textsuperscript{80}Sr could be a similar case, lying in a transitional region between axially prolate nuclei \textsuperscript{76,78}Sr and \textsuperscript{80}Zr, and axially oblate nuclei residing at and beyond \textsuperscript{84}Mo.

To investigate further, we compare to level energy systematics. The ratio of the $4^+_1$ and $2^+_1$ energies ($R_{42}$) in particular is well-established as a metric by which to distinguish rigid rotors ($R_{42}=3.33$) from vibrational ($R_{42}=2$) systems. This ratio is plotted against neutron number for Kr, Sr and Zr isotopes in Fig.~\ref{fig:PMMU_Calculations} (b) and for $N=40,42$ isotones against proton number in Fig.~\ref{fig:PMMU_Calculations} (c). Notably, no nuclei considered here reach the rotational limit, which can be explained by an enhancement in the ground-state binding due to proximity to the line of $N=Z$ (see e.g. Ref.~\cite{ref:Satula_97}). Nonetheless, \textsuperscript{80}Sr exhibits a consistently lower $R_{42}$ value than \textsuperscript{76,78}Sr and \textsuperscript{80}Zr, for which $R_{42}$ approaches a local maximum, and instead remains approximately consistent with its $N=42$ isotonic partners. The even-even isotonic neighbour of \textsuperscript{80}Sr, \textsuperscript{78}Kr, has been investigated thoroughly through Coulomb excitation~\cite{ref:Becker_06}, permitting a full analysis using the Kumar-Cline rotational invariants~\cite{Kumar72,Cline86}. These invariants allow for a model independent determination of $\left<\cos\left(3\gamma\right)\right>=0.41(6)$, indicating a significant role for triaxiality in that system. This is consistent with the observation for \textsuperscript{80}Sr made in the present work.  

\section{Conclusions}\label{Conclusions}

To conclude, we report on a sub-barrier Coulomb excitation measurement of \textsuperscript{80}Sr, performed at TRIUMF-ISAC. Despite large background present in this work, we are able to report the first measurement of $Q_s(2^+_1) = 0.45^{+0.83}_{-0.88}$ in \textsuperscript{80}Sr. Comparison with the expected value for a rigid rotor indicates a system which has not reached prolate axiality, and behaves consistently with its $N=42$ isotonic neighbours. A comparison was performed with calculations performed with the PMMU interaction which are able to include deformation-driving interactions between the key $1g_{9/2}2d_{5/2}$ orbitals, and which predict a $Q_s(2^+_1)$ value more consistent with an axially prolate system. These calculations, however, predict a shape change between the prolate and oblate minima to occur between \textsuperscript{80}Zr and \textsuperscript{84}Mo, with $N=42$ \textsuperscript{82}Zr a transitional point. One hypothesis would be that \textsuperscript{80}Sr might similarly lie in this transitional region. Importantly, with the present result inconsistent with axial prolate deformation at the level of more than $1.5\sigma$, it indicates that the anticipated island of strongly deformed axial systems around $N=Z=40$ is likely confined to \textsuperscript{76,78}Sr and \textsuperscript{78,80}Zr.

\section*{Acknowledgements}

The authors thank the beam-delivery team at TRIUMF-ISAC for providing the beams used in the present work, and J. Russell for a useful discussion during analysis. Work at the University of Surrey was supported under UKRI Future Leaders Fellowship grant no. MR/T022264/1. Work at the University of Surrey and University of York was supported by the Science and Technologies Facilities Council (STFC). This work has been supported by the Natural Sciences and Engineering Research Council of Canada (NSERC), The Canada Foundation for Innovation and the British Columbia Knowledge Development Fund. TRIUMF receives federal funding via a contribution agreement through the National Research Council of Canada. Supported by the U.S. Department of Energy, Office of Science, Office of Nuclear Physics under contract DE-AC02-06CH11357. Work at Lawrence Livermore National Laboratory was performed under the auspices of the U.S. Department of Energy under contract DE-AC52-07NA27344.

\bibliographystyle{model1-num-names}
\bibliography{bibliography}


\end{document}